\documentclass[aps,reprint,superscriptaddress]{revtex4}
\usepackage[english]{babel}
\usepackage{amssymb,amsmath}
\usepackage{graphicx}
\usepackage{multirow}

\begin{document}

\title{Atomic scale shot-noise using broadband scanning tunnelling microscopy}

\author{F. Massee}
\affiliation{Laboratoire de Physique des Solides (CNRS UMR 8502), B\^{a}timent 510, Universit\'{e} Paris-Sud/Universit\'{e} Paris-Saclay, 91405 Orsay, France}
\author{Q. Dong}
\affiliation{Centre de Nanosciences et de Nanotechnologies, CNRS, Universit\'{e} Paris-Sud, Universit\'{e} Paris-Saclay, C2N - Marcoussis, 91460 Marcoussis, France}
\author{A. Cavanna}
\affiliation{Centre de Nanosciences et de Nanotechnologies, CNRS, Universit\'{e} Paris-Sud, Universit\'{e} Paris-Saclay, C2N - Marcoussis, 91460 Marcoussis, France}
\author{Y. Jin}
\affiliation{Centre de Nanosciences et de Nanotechnologies, CNRS, Universit\'{e} Paris-Sud, Universit\'{e} Paris-Saclay, C2N - Marcoussis, 91460 Marcoussis, France}
\author{M. Aprili}
\affiliation{Laboratoire de Physique des Solides (CNRS UMR 8502), B\^{a}timent 510, Universit\'{e} Paris-Sud/Universit\'{e} Paris-Saclay, 91405 Orsay, France}
\email{freek.massee@u-psud.fr}
\date{\today}

\begin{abstract}
We have developed a broadband scanning tunnelling microscope capable of conventional, low frequency ($<$\thinspace10\thinspace kHz), microscopy as well spectroscopy and shot-noise detection at 1\thinspace MHz. After calibrating our AC circuit on a gold surface, we illustrate our capability to detect shot-noise at the atomic scale and at low currents ($<$\thinspace1\thinspace nA) by simultaneously measuring the atomically resolved differential conductance and shot-noise on the high temperature superconductor Bi$_{2}$Sr$_{2}$CaCu$_{2}$O$_{8+x}$. We further show our direct sensitivity to the temperature of the tunnelling electrons at low voltages. Our broadband probe opens up the possibility to study charge and correlation effects at the atomic scale in all materials accessible to STM.
\end{abstract}

\maketitle

\section{Introduction}
A direct consequence of the discreteness of the electron charge is the presence of time-dependent fluctuations of the electronic current, called shot-noise. In the tunnelling limit, the power spectral density of the shot-noise is given by $S_{I} = 2q|I|F$, were $q$ is the charge of the tunnelling entity, $I$ the average current and $F$ the Fano factor that encodes correlation effects (see \cite{blanter_pr_2000} for a review of the theory and application of shot-noise measuments in mesoscopic systems). Particularly suited for mesoscopic systems, shot-noise has been a powerful tool in the study of electron-electron correlations in e.g. Kondo systems \cite{meir_prl_2002, delattre_naturephysics_2009, yamauchi_prl_2011, ferrier_naturephysics_2016}, Coulomb blockade scenarios \cite{cleland_prl_1990, altimiras_prl_2014}, and has been used to reveal fractional charges in the fractional quantum hall regime \cite{depicciotto_nature_1997, bid_prl_2009}, as well as doubling of the charge transferred for Andreev tunnelling \cite{jehl_nature_2000, ronan_pnas_2016}. The spatial resolution of these experiments is typically determined by the width of the contact leads to the sample and therefore ranges from tens of nanometers to several microns. This makes it impossible to study single impurities in a host material, for instance single atom Kondo physics, or correlation effects at impurities in doped Mott insulators such as the high-Tc cuprates. Previous efforts have shown the feasibility to increase the spatial resolution of shot-noise detection by using a scanning tunnelling microscope (STM) \cite{birk_prl_1995, kemiktarak_nature_2007, herz_nanoscale_2013, burtzlaff_prl_2015}, allowing for the study of junctions formed with nanometer sized metallic particles \cite{birk_prl_1995}. This is because the shot-noise is a current noise and will consequentially have the same atomically resolved resolution as the tunnelling current in an STM. Most of these approaches relied on relatively high tunnelling currents \cite{kemiktarak_nature_2007, herz_nanoscale_2013, burtzlaff_prl_2015}, which due to the high local electric fields result in unstable junctions with many (weakly Van der Waals bound) correlated electron systems.

Here we demonstrate our ability to measure shot-noise at the atomic scale at low currents ($<$\thinspace1\thinspace nA) and high junction resistances ($>$\thinspace10\thinspace M$\Omega$) simultaneously with DC tunnelling spectroscopy. We achieve this by combining a home-built low temperature scanning tunnelling microscope with an LC$_{\text{cable}}$ resonant circuit, and a cryogenic amplifier based on a high electron mobility transitor (HEMT) \cite{dong_apl_2014}. First, we describe the circuit in detail and present shot-noise measurements on a gold surface, which serves as a calibration of the circuit. We then show the sensitivity of shot-noise to the electron temperature and our ability to measure differential conductance at MHz modulation frequencies. We illustrate the capabilities of our atomic scale shot-noise microscope by simultaneously measuring the atomically resolved differential conductance and shot-noise on the high temperature superconductor Bi$_{2}$Sr$_{2}$CaCu$_{2}$O$_{8+x}$.

\begin{figure}[h]
	\centering
	\includegraphics[width=15cm]{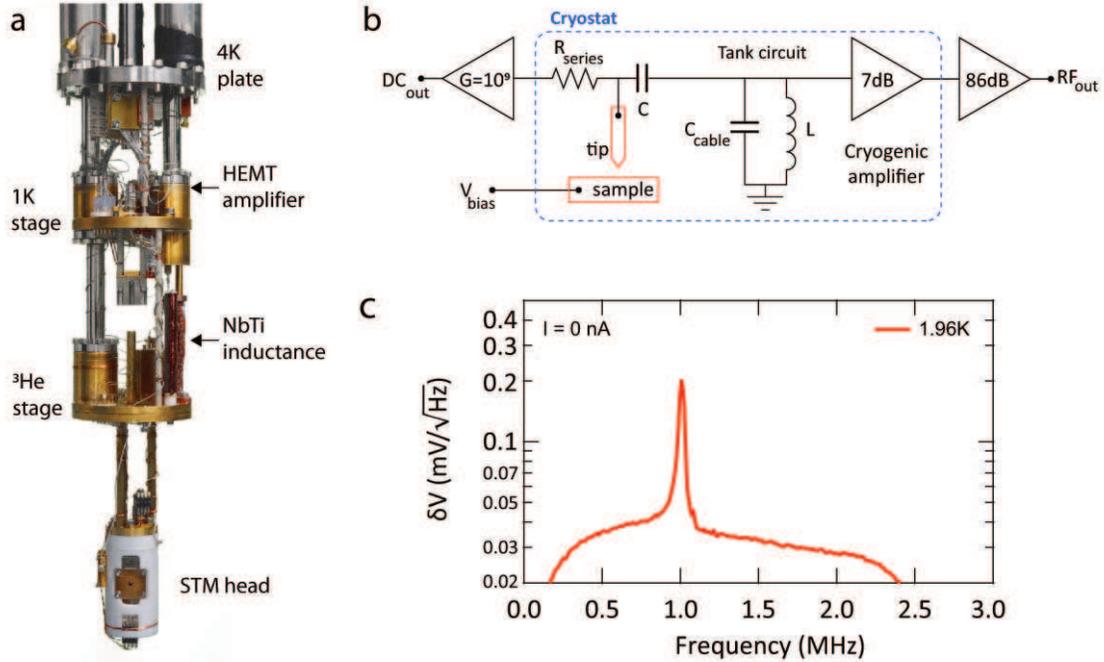}
	\caption{\label{fig:broadband_stm} Broadband STM. a) Home-built scanning tunnelling microscope (design based on \cite{pan_revsciinstrum_1999}) outfitted with broadband circuitry. b) In parallel to a conventional junction and its stray capacitance, R$_{\text{J}}$ and C$_{\text{J}}$ respectively, a finite frequency circuit is added consisting of an LC$_{\text{cable}}$ resonance and a HEMT amplifier. The inductance is a coil of superconducting NbTi wire, with L = 155\thinspace$\mu$H, which gives a resonance at 1 MHz for a cable capacitance of C$_{\text{cable}} \sim$ 160\thinspace pF. A C = 6.8\thinspace nF capacitance separates the DC and AC circuits and a resistance, R$_{\text{series}}$ = 400\thinspace k$\Omega$, in series with the tunnel junction suppresses cross-talk and ensures the LC$_{\text{cable}}$ resonance has a proper ground. A bandpass filter further removes the DC and high frequency components before readout with a spectrum analyser. c) Thermal noise of the LC$_{\text{cable}}$ resonance at T = 2\thinspace K.}
\end{figure}

\section{The circuit}

At low frequencies ($<$\thinspace 100\thinspace kHz) the noise spectral density is dominated by 1/f noise and mechanical resonances. At frequencies $>$\thinspace 100\thinspace kHz where shot-noise can be determined accurately, voltage noise from commonly used (high input impedance) room temperature current amplifiers is converted into current noise by the cable capacitances. As this spurious current noise grows linearly with frequency it rapidly overwhelms the shot-noise. To overcome the impedance mismatch at low temperature and loss through cable capacitances, we have equipped our home-built scanning tunnelling microscope (Fig. \ref{fig:broadband_stm}a) with an LC$_{\text{cable}}$ resonant circuit with its resonance tuned in the low MHz regime, followed by a low-noise high electron mobility transistor \cite{dong_apl_2014} (Fig. \ref{fig:broadband_stm}b,c). A capacitance (C = 6.8\thinspace nF) separates the current line from the LC$_{\text{cable}}$ resonant circuit, enabling determination of the shot-noise simultaneous with conventional STM measurements. The current noise is converted to voltage noise at the HEMT input by the modulus of the impedance at resonance, $|Z_{\text{res}}|$. This impedance is set by the quality factor of the LC$_{\text{cable}}$ resonance and the resistance, R$_{\text{series}}$ = 400\thinspace k$\Omega$, in the current line that suppresses cross-talk between the two circuits and ensures the LC$_{\text{cable}}$ resonance has a proper ground.

As the conversion of current noise to voltage noise goes with the square of the impedance at resonance, whereas the Johnson-Nyquist (i.e. thermal) noise is linear in impedance, a higher impedances at resonance (i.e. higher quality factor) is beneficial. In order to have a relatively high impedance at resonance, yet still small compared to junction resistances of $>$\thinspace10\thinspace M$\Omega$ typically used for studying correlated electron systems, we use a home-wound inductance of 155\thinspace $\mu$H made from Cu-clad NbTi wire which has an impedance at resonance of 170\thinspace k$\Omega$ (see below). The AC voltage noise output from the HEMT is further amplified at room temperature followed by a bandpass filter to remove 1/f and high frequency noise. The voltage noise at the spectrum analyser can then be approximated by:
\begin{equation}\label{equation:noise}
S_{tot} = (2q|I|F|Z_{\text{res}}|^2 + 4k_{\text{B}}T|Z_{\text{res}}| + \delta V_{\text{amp}}^2)G^2,
\end{equation}
where $G^2$ is the total gain of the amplification chain, $\delta V_{\text{amp}}$ is the combined input and output voltage noise of the various amplifiers and T is the temperature of the LC$_{\text{cable}}$ resonator, which we show to be identical to the temperature of the STM (T$_{\text{stm}}$) and of the tunnel junction (T$_{\text{e}}$).

\section{Calibration}

\begin{figure}[h]
	\centering
	\includegraphics[width=15cm]{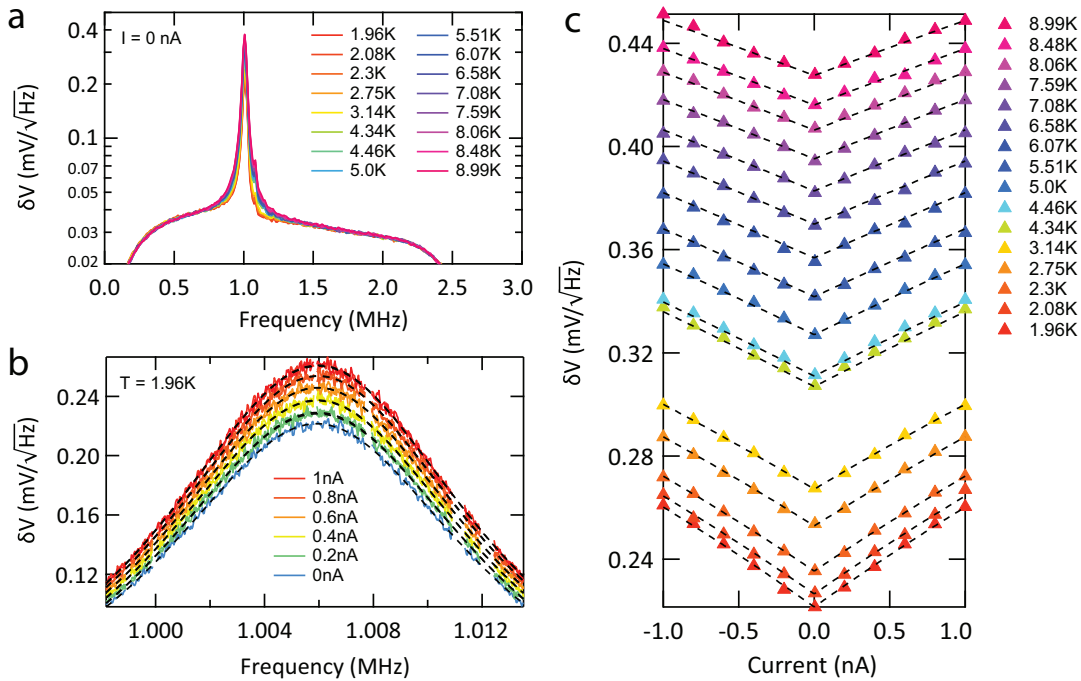}
	\caption{\label{fig:circuit} Circuit calibration. a) Thermal noise as function of frequency for different temperatures. b) Current dependence of the total noise at 2\thinspace K at a fixed junction resistance of 100\thinspace M$\Omega$, dotted lines are Lorentzian fits to extract the peak amplitude. c) Magnitude of the amplitude spectral density as function of current for a range of temperatures, dashed lines are fits using Eq. (\ref{equation:noise}) with $|Z_{\text{res}}|$ = 170\thinspace k$\Omega$, G = 4.5x10$^{4}$ (G$_{\text{HEMT}}$ = 2.25) and $\delta$V$_{amp}$ = 2.42\thinspace nV/$\sqrt{\text{Hz}}$.}
\end{figure}

To calibrate our circuit we bring a Pt/Ir tip in tunnelling contact with a gold sample. The shot-noise in this case is purely Poissonian and the charge single electrons, i.e. F = 1 and q = e. In order to accurately determine the impedance at resonance, as well as the gain and the noise of the amplification chain, we measure the shot-noise of the junction for a range of temperatures between 1\thinspace K and 10\thinspace K. We use a sufficiently high junction resistance ($>$\thinspace10\thinspace M$\Omega$) to not be affected by thermal effects. We further assume that the quality factor of the resonant circuit (Q $\sim$ 25), which is at the same temperature as the sample and tip, is temperature independent below the superconducting transition temperature of NbTi, T$_{\text{c}}$ $\sim$ 10\thinspace K. The temperature of the HEMT, which is mounted on a different stage, is kept below 5\thinspace K, where its input noise and gain can safely be assumed to be temperature independent \cite{dong_apl_2014}. Figure \ref{fig:circuit}a shows the noise as function of temperature in absence of a tunnelling current, i.e. the thermal noise of the resonance plus the background noise of the amplification chain. Upon establishing a current, additional noise is observed as can be seen for T $\sim$ 2\thinspace K in Fig. \ref{fig:circuit}b, which is the shot-noise we are interested in. We fit the peak of the resonance with a Lorentzian (dotted lines in Fig. \ref{fig:circuit}b) to extract its maximum value. Figure \ref{fig:circuit}c shows the peak amplitudes of the total noise as function of tunnelling current for a range of temperatures - the value at I = 0\thinspace nA corresponds to the thermal noise plus the background from the amplification chain as shown in Fig. \ref{fig:circuit}a. To extract all relevant circuit parameters we finally fit the current and temperature dependence of Fig. \ref{fig:circuit}c using Eq. (\ref{equation:noise}), setting F = 1, giving: $|Z_{\text{res}}|$ = 170\thinspace k$\Omega$, G = 4.5x10$^{4}$ (of which the gain of the HEMT, G$_{\text{HEMT}}$, is 2.25) and $\delta$V$_{\text{amp}}$ = 2.42\thinspace nV/$\sqrt{\text{Hz}}$. Subsequent cooldowns have given similar results. 

\section{Electron temperature}

\begin{figure}[h]
	\centering
	\includegraphics[width=10cm]{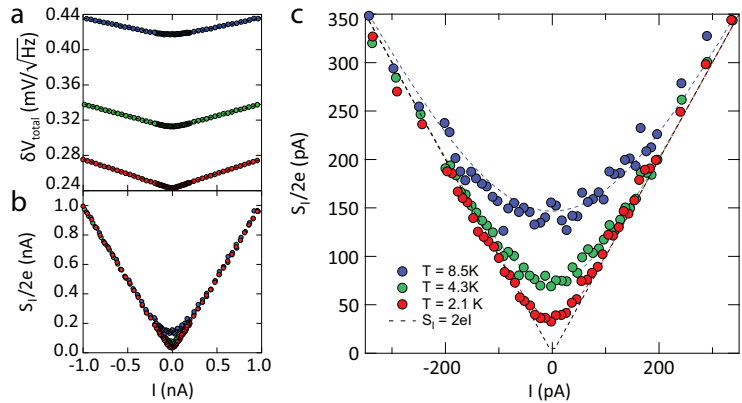}
	\caption{\label{fig:thermally_limited_noise} a) Total voltage noise at three different temperatures for the same junction resistance of 10\thinspace M$\Omega$. b) Shot-noise after taking into account the thermal noise of the circuit (but not of the tunnel junction) and the noise of the amplification chain, see Eq. (\ref{equation:noise}). c) Enlargement of b) corresponding to energies where the thermal noise of the junction becomes limiting (dots: measurement, coloured dashed: fits with T$_{\text{fit}}$ = T$_{\text{stm}}$, black dashed: $S_I=2e|I|$).}
\end{figure}

An additional benefit of our broadband scanning tunnelling microscope is that through shot-noise we are directly sensitive to the temperature of the tunnelling electrons. At sufficiently low voltages, the shot-noise will become thermally limited as can be appreciated from the full expression of the shot-noise, $S_{I} = 2q|I|\text{coth}(eU/2k_{B}T_{\text{e}})F$, where T$_{\text{e}}$ is the electron temperature and U the bias voltage. This temperature need not necessarily be identical to the bath temperature that is used in the fits to determine the circuit components (Fig. \ref{fig:circuit}), as atomic scale Joule heating, or limited filtering and/or poor thermal anchoring could lead to a higher junction temperature. In Fig. \ref{fig:thermally_limited_noise} we show shot-noise measured at R$_{\text{J}}$ = 10\thinspace M$\Omega$ (E$_{\text{bias}}$ = $\pm$\thinspace 10mV) for three different temperatures. Since at these relatively low voltages and high currents the voltage drop over the series resistance in the current line (R$_{\text{series}}$ = 400\thinspace k$\Omega$) becomes non-negligible, we have taken into account that the voltage drop across the sample is reduced by a factor IR$_{\text{series}}$. As expected, once the voltage approaches the thermal energy of the electrons, eU $\sim$ 2k$_{\text{B}}$T, the noise levels off. The excellent match between the data and the fits (dotted lines) using T$_{\text{e}}$ = T$_{\text{stm}}$ shows that our tip, sample and the resonator are properly thermalised to the bath, and that at $<$\thinspace 1\thinspace nA current Joule heating is negligible. Conversely, it proves we can use shot-noise as an accurate thermometer of the tunnel junction.

\section{Atomic resolution and MHz modulation}

\begin{figure}[h]
	\centering
	\includegraphics[width=10cm]{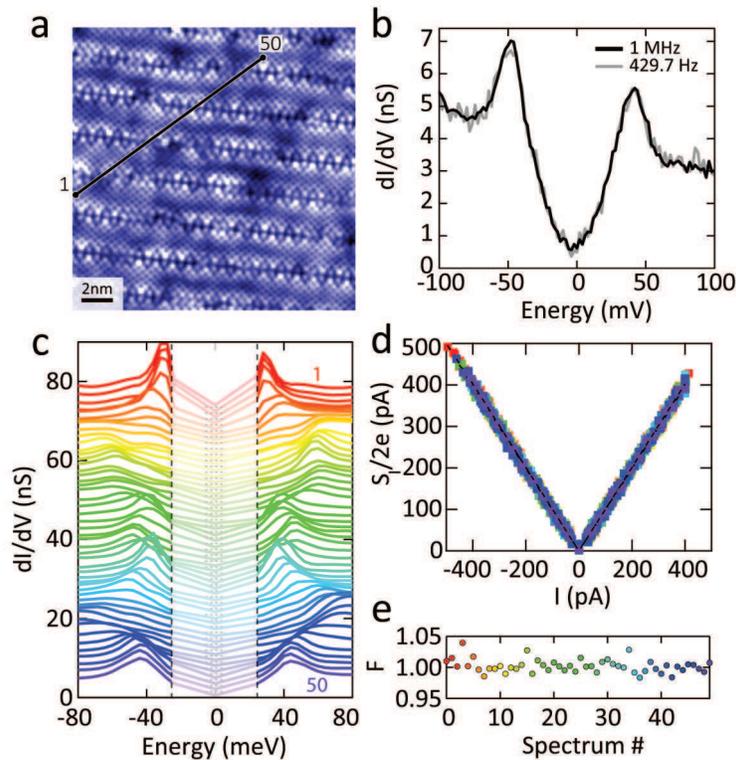}
	\caption{\label{fig:bi2212} a) Atomic resolution image of Bi$_{2}$Sr$_{2}$CaCu$_{2}$O$_{8+x}$ taken with our broadband STM. b) Differential conductance using a voltage modulation of 1mV at 429.7\thinspace Hz (DC circuit) and 1\thinspace MHz (AC circuit) with identical averaging times per point, taken simultaneously with each other. c) Spatial variation of differential conductance spectra taken along the line indicated in a), focussing on $|E|>30$\thinspace mV. d) Shot-noise measurements taken simultaneously with the spectra in c), normalised using the circuit parameters determined in Fig. \ref{fig:circuit}. e) Fano factor extracted from the slope of the shot-noise measurements in d). Simultaneously taken measurements in c-e have the same colour.}
\end{figure}

We now proceed to demonstrate that we can atomically resolve spectroscopy and simultaneous shot-noise measurements, and show the benefits of measuring differential conductance at finite frequency, by turning our attention to the high temperature superconductor Bi$_{2}$Sr$_{2}$CaCu$_{2}$O$_{8+x}$ (Bi2212). Figure \ref{fig:bi2212}a displays the atomically resolved lattice of an optimally doped sample (T$_{\text{c}}$ $\sim$ 90\thinspace K) with its characteristic incommensurate super-modulation. Using a conventional low frequency (429.7\thinspace Hz) lock-in technique, we measure the differential conductance shown in Fig. \ref{fig:bi2212}b. An additional voltage modulation with a frequency corresponding to the resonance frequency of our LC$_{\text{cable}}$ circuit can be added to simultaneously measure the differential conductance at 1\thinspace MHz. As is apparent from Fig. \ref{fig:bi2212}b, the elevated frequency of the lock-in detection enhances the signal-to-noise significantly, reducing the required measurement time for spectroscopic measurements. 

To measure the shot-noise simultaneously with the differential conductance as function of spatial location, we reserve our MHz circuit for shot-noise detection, while applying a low frequency voltage modulation to determine the differential conductance. Figure \ref{fig:bi2212}c shows a series of spectra taken at various locations including regions of small ($\Delta_{\text{peak-peak}} \sim$ 30\thinspace mV) to big ($\Delta_{\text{peak-peak}} \sim$ 70\thinspace mV) gaps. All spectra are taken with a setup voltage and current of 80\thinspace mV and 400\thinspace pA respectively. As the feedback loop is turned off and the voltage is swept from 80\thinspace mV to -80\thinspace mV, the current reduces to zero and changes sign, allowing us to simultaneously measure the current dependence of the shot-noise. The current in this case is predominantly from states outside the gap, i.e. $|E|>$ 30\thinspace mV, and the corresponding shot-noise is presented in Fig. \ref{fig:bi2212}d. Despite the large variation in gap size of the various locations, the states outside the gap that we address here show little to no variation in their shot-noise - the slope of the noise, or Fano factor, plotted in Fig. \ref{fig:bi2212}e is equal to one for all spectra with an error bar of a few percent. This means that outside the gap the quasi-particles responsible for tunnelling have a charge equal to a single electron charge, e, and tunnelling events are uncorrelated within our experimental error, F = 1.

\section{Discussion}
We have shown our ability to resolve shot-noise at the atomic scale simultaneously with the conventional, DC, operation of a scanning tunnelling microscope at low currents and high junction resistances with an experimental uncertainty of a few percent in the slope of the noise. One of the crucial ingredients in our circuit is the use of an impedance matching cryogenic amplifier. Due to its very low input noise at low temperature ($\delta$V$_{\text{input}}$ = 0.22\thinspace nV/$\sqrt{\text{Hz}}$ at 1\thinspace MHz), our cryoHEMT allows us to detect shot-noise with a high signal-to-noise ratio at low currents. For comparison, a low frequency circuit ($\sim$\thinspace 40\thinspace kHz) with room temperature amplification allowed us to detect signs of shot-noise after averaging each data point for several minutes at currents up to 1\thinspace nA. The system described in this work achieves the same signal-to-noise for sub-second averaging times. This can be further improved upon by suppressing the thermal noise at even lower temperature, and/or by increasing the impedance at resonance.

In order to address changes in the Fano factor with an even higher accuracy than a few percent reported here, and/or for low junction resistances, i.e. when R$_{\text{J}} >>$ Z$_{\text{res}}$ does not hold, one important factor needs to be taken into account that was negligible in the measurements we have presented thus far. The voltage to ground at the input of the cryogenic amplifier is proportional to the impedance at resonance of the LC$_{\text{cable}}$ circuit parallel to R$_{\text{series}}$ \textit{and} R$_{\text{J}}$. As the latter reduces, so does the absolute value of the detected noise as $Z_{\text{res}}$ in Eq. (\ref{equation:noise}) should be replaced by the transimpedance $Z_{0} = Z_{\text{res}}R_{\text{J}} / (Z_{\text{res}} + R_{\text{J}})$ \footnote{We can safely ignore the capacitance, C, since $Z_c=1/(2\pi f C) \sim 25 \thinspace \Omega$ is negligible.}. If the current-voltage characteristic is non-linear, as is the case for example for Bi2212, the dynamical resistance has to be substituted for R$_{\text{J}}$ in this correction.

The atomic scale detection of shot-noise we present here opens up a new avenue to study the dynamics of tunnelling and correlation effects around single impurities in all systems accessible by scanning tunnelling microscopy. 

\textit{Note: a similar approach for atomic scale noise detection is reported by Bastiaans et al.: ``Amplifier for scanning tunneling microscopy at MHz frequencies''.}

We thank M. P. Allan, J. Gabelli and F. Pierre for useful discussions, and Y. K. Huang and M. S. Golden for providing the Bi$_{2}$Sr$_{2}$CaCu$_{2}$O$_{8+x}$ samples used in this study. FM would like to acknowledge funding from H2020 Marie Sk\l{}odowska-Curie Actions (grant number 659247) and the ANR (ANR-16-ACHN-0018-01).

\end{document}